# Caustic graphene plasmons with Kelvin angle

Xihang Shi[1], Xiao Lin[1], Fei Gao[1], Hongyi Xu[1], Zhaoju Yang[1], Baile Zhang[1,2,*]

1. *Division of Physics and Applied Physics, School of Physical and Mathematical Sciences, Nanyang Technological University, Singapore 637371, Singapore*
2. *Centre for Disruptive Photonic Technologies, Nanyang Technological University, Singapore 637371, Singapore*

**Electronic mail of corresponding author: blzhang@ntu.edu.sg*

## Abstract

A century-long argument made by Lord Kelvin that all swimming objects have an effective Mach number of 3, corresponding to the Kelvin angle of 19.5 degree for ship waves, has been recently challenged with the conclusion that the Kelvin angle should gradually transit to the Mach angle as the ship's velocity increases. Here we show that a similar phenomenon can happen for graphene plasmons. By analyzing the caustic wave pattern of graphene plasmons stimulated by a swift charged particle moving uniformly above graphene, we show that at low velocities of the charged particle, the caustics of graphene plasmons form the Kelvin angle. At large velocities of the particle, the caustics disappear and the effective semi-angle of the wave pattern approaches the Mach angle. Our study introduces caustic wave theory to the field of graphene plasmonics, and reveals a novel physical picture of graphene plasmon excitation during electron energy-loss spectroscopy measurement.

Graphene plasmons [1-5], as typical two-dimensional (2D) plasmons confined on a 2D surface [6], have recently emerged as a new research branch in photonics that attracts substantial research efforts. A typical quantitative measurement on graphene plasmons relies on electron energy-loss spectroscopy (EELS) measurement [7-10]. In studying planar surfaces with EELS, a particularly instructive situation is that when the electron is directed parallel to the surface [11]. This configuration has been extensively studied for other materials both experimentally and theoretically before the discovery of graphene [12-19], and has also been studied for graphene [20-22] recently. The unique dispersion of graphene plasmons compared to surface plasmons at a metal/dielectric interface may cause interesting phenomena that have not be revealed.

The fact that 2D plasmons, including graphene plasmons, exhibit a dispersion similar to that of deep-water waves in the long wavelength limit, has already been known for more than fifty years [1,5,6,23]. This implies that many deep-water-wave phenomena can find counterparts in graphene plasmons. Recently, a century-long celebrated prediction made by Lord Kelvin in 1880s that the semi-angle of ship waves on water surface is fixed at $19.5°$, independent of the ship's velocity [24,25], has been challenged by the conclusion that the ship-wave semi-angle should gradually transit from the Kelvin angle to the Mach angle as the ship's velocity increases [26,27]. Kelvin made his prediction based on the dispersion of deep-water waves, while the recent studies argued that the deep-water waves excited by a ship should have preferred directions determined by the Frouder number, or the velocity of the ship, given a fixed ship size.

In this Letter, we adopt approaches of caustic wave theory to analyze the graphene plasmons excited by a swift charged particle moving above the graphene plane.

Previously, caustics have been treated predominantly on the level as envelopes of families of rays [28]. Recently, they have been used to generate arbitrary bending surface plasmon beams [29] on a metal surface, and discussed in the scattering of electron flow from graphene *p*-*n* junctions [30,31]. Yet caustics have not been discussed in the context of graphene plasmons. Here we study the wave aspects of caustics formed by graphene plasmons when a swift charged particle is moving parallel to the graphene plane. We find that at a relatively small velocity ($\sim 0.1c$ or smaller, where $c$ is the velocity of light in vacuum) of the swift charged particle, the stimulated graphene plasmons are confined within the nonsingular caustic boundaries with the semi-angle of $19.5°$, i.e. the Kelvin angle. These caustic boundaries are associated with cusps wave fronts and thus can be classified as fold caustics [28]. Each point within the caustic boundaries is covered by rays twice, whereas outside there is no ray. The more exact calculation with Airy integral distinguishes the caustic shadow from the caustic zone, thus provides a complete wave description of caustic graphene plasmons. As the velocity of the charged particle increases, the graphene-plasmonic waves reaching the caustic boundaries become weak, which blurs the caustics and eventually makes them disappear. The calculation shows the effective semi-angle of the wave pattern approaches the Mach angle, being similar to the recent studies of ship waves in fluid mechanics.

We first introduce the calculation model, as illustrated in Fig. 1, where a particle with charge $q$ moves along $\hat{z}$ direction with a uniform velocity $v$ parallel to an isolated graphene sheet at $y = d$. The current density that this charged particle generates is

$$\mathbf{J}(\mathbf{r},t) = \hat{z}qv\delta(x)\delta(y)\delta(z-vt), \tag{1}$$

and its Fourier transform is

$$\mathbf{J}(\mathbf{r},\omega)=\hat{z}\frac{q}{4\pi^2\rho}e^{i\omega z/v}\delta(\rho), \tag{2}$$

where $\rho=\sqrt{x^2+y^2}$. The evanescent fields from the swift charged particle can excite graphene plasmons on graphene. In the calculation we set $d=1\mu\text{m}$. The isolated graphene is assumed to have chemical potential $\mu_c=0.15\text{eV}$, and scattering rate $\Gamma=0.11\text{meV}$ at the room temperature $T=300\text{K}$ [32]. The frequency dependent complex conductivity $\sigma(\omega)$ of the isolated graphene is computed from Kubo formula [32,33]. Each frequency component of the graphene plasmons is exactly derived by taking the residue of Sommerfeld pole [34]. The frequency components of excited graphene plasmons are mostly below $50\text{THz}$, above which the interband transition in graphene will dominant and cause large loss [1,2,23]. We use the vertical component of electrical field $E_y$ to represent the transverse-magnetic (TM) graphene plasmons

$$E_y(\omega)=\frac{-iq}{4\pi v\varepsilon_0}\frac{k_y^2}{k_x}e^{ik_yd+ik_xx+ik_zz}, \tag{3}$$

where $\varepsilon_0$ is the vacuum permittivity, $k_y=-2\omega\varepsilon_0/\sigma(\omega)$ indicates the confinement of graphene plasmons, and $k_z=\omega/v$ and $k_x=\sqrt{\omega^2\varepsilon_0\mu-k_y^2-k_z^2}$ are the wave vectors of graphene plasmons. The field distribution at time $t$ is the Fourier integral of Eq. (3)

$$\begin{aligned}E_y(y=d,t)&=2\,\text{Re}\left[\int\frac{-iq}{4\pi v\varepsilon_0}\frac{k_y^2}{k_x}e^{ik_yd}e^{ik_xx+ik_zz-i\omega t}d\omega\right],\\&=2\,\text{Re}\left[\int e_y(\omega)e^{ik_yd}e^{i\psi(\omega)t}d\omega\right]\end{aligned} \tag{4}$$

where $\psi(\omega)=k_x\frac{x}{t}+k_z\frac{z}{t}-\omega$.

We then discuss the situation when the velocity of the particle is set to be $v=0.1c$. We numerically carry out the integration in Eq. (4) to get the wave pattern, as shown in

Fig. 2a. The left part is the absolute value of the electrical field $|E_y(r,t)|$ and the right part is the absolute value of the total electrical field $|E_{tot}(r,t)|$. The arrow indicates the position of the charged particle. The semi-angle of 19.5°, i.e. the Kelvin angle, is clearly seen. A plane-like wave is inside the wave pattern, following the charged particle.

An intuitive picture, following the approach of Kelvin [24,25], from the viewpoint of ray theory, is shown in Fig. 2b. Assuming the particle moves from point A to point B with velocity $v$. The ship waves of graphene plasmons excited at point A will propagate in all directions with different frequencies. In the propagation direction of angle $\theta$ with respect to the trajectory of the particle, the phase velocity $v_p$ of graphene plasmons must satisfy the stationary condition $v_p = v\cos\theta(\omega)$ [24,25]. When the particle arrives at point B, The dashed red circle represents the locus of all the arrived phases of graphene plasmon waves. However, the group velocity is $v_g = v_p\, \beta^2/(2\beta^2 - k_0^2)$ [35]. It can be shown that, when $v = 0.1c$, $v_g \cong v_p/2$ [35], and thus the locus of the energy of arrived waves form the solid red circle with the diameter only half of the red dashed one. The waves propagating in the direction $\theta = 35°$ form the Kelvin caustic boundary with angle $\alpha = \sin^{-1}(1/3) = 19.5°$, as shown by the solid black lines in Figs. 2b and c. According to ray tracing theory [35], the wave fronts are drawn as the blue lines in Fig. 2c. All the wave fronts have cusps on the caustic boundary.

We can asymptotically evaluate the integral in Eq. (4) with the stationary phase method. The integration contour of Eq. (4) can be deformed in the complex $\omega$ plane and the only contribution comes from the path of steepest descent [24] around the stationary value $\omega = \omega_s$, where

$$\frac{d\psi(\omega)}{d\omega} = \frac{dk_x}{d\omega}\frac{x}{t} + \frac{dk_z}{d\omega}\frac{z}{t} - 1 = 0. \tag{5}$$

Equation (5) is equivalent to the group velocity concept in the ray tracing theory [35]. The expansion of $\psi$ around $\omega_s$ is

$$\psi(\omega) = \psi(\omega_s) + \frac{1}{2}\psi''(\omega_s)(\omega - \omega_s)^2 + O\left(|\omega - \omega_s|^3\right) \tag{6}$$

The integral in Eq. (4) can be approximated as

$$E_y(y = d, t) = 2\,\mathrm{Re}\left[e_y(\omega_s)\frac{\sqrt{2\pi}}{\sqrt{\psi''(\omega_s)t}}e^{ik_y(\omega_s)d + i\psi(\omega_s)t + i\pi/4}\right]. \tag{7}$$

Equation (7) diverges when $\psi''(\omega) = 0$. In this case the path of steepest descent has to be chosen differently. It can be shown that for the frequency component $\omega = \omega_c$, where

$$\frac{d^2\psi(\omega)}{d\omega^2} = \frac{d^2k_x}{d\omega^2} = 0, \tag{8}$$

the first derivative of group velocity with respect to frequency vanishes. Therefore the rays run together near such points, and the locus of such points form caustics or caustic boundary, which separates a region without rays from a region covered by rays twice. We get the caustic boundary equation by combining Eq. (5) and Eq. (8)

$$2\sqrt{2}\sqrt{1 - \left(\frac{v}{c}\right)^2}\frac{x}{vt} + \frac{z}{vt} = 1, \tag{9}$$

whose solution is shown by the black solid line in Fig. 2c. When $v = 0.1c$, the exact angle of the caustics is $\alpha = 19.56°$.

To calculate the field around the caustics, we expand $\psi$ around $\omega_c$

$$\psi(\omega)=\psi(\omega_c)+\psi'(\omega_c)(\omega-\omega_c)+\frac{1}{6}\psi'''(\omega_c)(\omega-\omega_c)^3+O\left((\omega-\omega_c)^4\right). \quad (10)$$

In the neighborhood of the caustics, the first derivative of phase (i.e., $\psi'(\omega_c)$) is almost, but not exactly, equal to zero. Substituting Eq. (10) into Eq. (4), the asymptotic form of $E_y(y,t)$ can be calculated with Airy integral [35],

$$E_y(y=d,t)=2\,\mathrm{Re}\left\{2\pi e_y(\omega_c)\left[\frac{t}{2}\psi'''(\omega_c)\right]^{-\frac{1}{3}}e^{ik_y(\omega_c)d+i\psi(\omega_c)t}\mathrm{Ai}(\mathrm{X})\right\}. \quad (11)$$

where $\mathrm{Ai}(\mathrm{X})$ is the Airy integral function and

$$\mathrm{X}=\frac{t\psi'(\omega_c)}{\left(t\psi'''(\omega_c)/2\right)^{1/3}}. \quad (12)$$

When $\mathrm{X}=0$, Eq. (12) represents the caustic boundary. When $\mathrm{X}>0$, Eq. (12) represents the caustic shadow region, where, because of the Airy function, the field decays exponentially. When $\mathrm{X}=0.66$, Airy function reach $1/e$ of the its maximum value. We set $\mathrm{X}=0.66$ to represent the boundary of the caustic shadow, as shown by the dashed black line in Fig. 2c. The Airy function reaches maximum at $\mathrm{X}=-1.02$. The region between $\mathrm{X}=0$ and $\mathrm{X}=-1.02$ corresponds to the caustic zone. The field in the caustic zone is relatively strong in the neighborhood of caustics as the wave-field focusing effect on caustics [28]. The field focusing in the caustic zone can be observed in Fig. 2a near the caustic boundary. The Airy function vanishes at $\mathrm{X}=-2.34$. It means the field distribution has a minimum value, as shown by the dashed red line in Fig. 2c. It can also be seen in Fig. 2a that a dark line (relatively weak field amplitude) separates the neighborhood of caustics from the waves inside the caustics. The field amplitude in the neighborhood of the caustic decays more slowly, with $t^{-1/3}$ as shown by Eq. (11), than the field amplitude in the region inside the caustics, with $t^{-1/2}$ as shown by Eq. (7).

When the velocity of the charged particle is chosen to be $0.3c$, the wave pattern of graphene plasmons is shown in Fig. 3a. The solid white line represents the caustic boundary defined by Eq. (9). The dashed white line is the boundary of the caustic shadow defined by Eq. (12) with $\mathrm{X} = 0.66$ . Compared with Fig. 2a, the angle between the caustic boundary and the z axis increases slightly and the caustic shadow expands, but the lightest branch of the waves contracts. Besides, there is no plane-like wave at the center of the wave pattern.

To understand the transition of wave pattern, we rewrite Eq. (4) as an integration over propagation angle $\theta$,

$$\begin{aligned} E_y(y=d,t) &= 2\,\mathrm{Re}\left[\int e_y e^{ik_y d} e^{i\psi(\omega)t} \frac{d\omega}{d\theta} d\theta\right] \\ &= \mathrm{Re}\left[\int E_y(\theta) e^{i\psi(\omega)t} d\theta\right] \end{aligned}, \tag{13}$$

where

$$E_y(\theta) = 2 e_y e^{ik_y d} \frac{d\omega}{d\theta} . \tag{14}$$

The dependence of $|E_y(\theta)|$ on angle $\theta$ is shown in Fig. 3b with different velocities. When the velocity is $0.1c$, $|E_y(\theta)|$ has relatively high value from 0° to 60°. When the velocity increases to $0.3c$, $0.5c$ and $0.7c$, relatively high values of $|E_y(\theta)|$ occur at large propagation angles. The propagation angles at which the waves will reach the caustic boundary are $\theta = 35.1°$, $34.0°$, $31.5°$ and $26.8°$, correspondinging to velocities of $0.1c$, $0.3c$, $0.5c$ and $0.7c$, respectively, as marked by arrows in Fig. 3b at the horizontal axis.

However, the wave amplitudes at these angles for velocities $0.3c$, $0.5c$ and $0.7c$ are so weak that they are difficult to perceive. Therefore, as the velocity increases from $0.1c$, the caustics get blurred and eventually disappear. At this moment, the effective boundary of the wave pattern is determined instead by the outmost perceivable fields.

As shown in Fig. 2b and Fig. 3b, the outmost perceivable fields at large velocities of the particle are contributed by waves with large propagation angle $\theta$, corresponding to group velocity $v_g'$. It can be shown that as the particle's velocity $v$ increases, $v_g'$ tends to be independent of $v$ [35]. Therefore, $\alpha \approx \frac{v_g'}{v} \propto 1/v$ and the effective semi-angle of the wave pattern behaves like Mack angle. The wave patterns of graphene plasmons excited by the swift charged particle with velocities $0.5c$ and $0.7c$ are shown in Figs. 4a and b, respectively. Compared with Fig. 2a and Fig. 3a, the wave patterns get more contracted as the velocity further increases. In Fig. 2a, we find that in the horizontal line $z = 0$, the absolute of the total electrical field at the caustic point is 0.61 relative to its maximum value along the line. We use this as a criterion to determine the angle of the ship waves at different velocities of the charged particle, and the results are shown in Fig. 4c. The angle of $\alpha = 19.47°$ is the angle of the caustic boundary at zero velocity limitation from Eq. (9). We plot the Mach angle line $\alpha = 4.52/v$ with the coefficient 4.52 adopted from the wave pattern when the particle's velocity is 0.9c. The results show clearly the transition from Kelvin angle of $19.5°$ at small velocities of the charged particle to the Mach angle at large velocities, being similar to the recent studies in fluid mechanics [26,27].

In conclusion, we reveal a novel wave phenomenon of graphene plasmons that is relevant to EELS measurement. We find that graphene plasmons excited by a swift charged particle moving above graphene can form a caustic wave pattern with semi-angle equal to the Kelvin angle, when the velocity of the charged particle is slow. At large velocities, the effective semi-angle of graphene plasmons approach the Mach angle. These results incorporate the recent development in fluid mechanics into the understanding of graphene plasmon excitation in EELS measurement, and introduce caustic wave theory into the field of graphene plasmonics, which might find use in the manipulation of graphene plasmons.

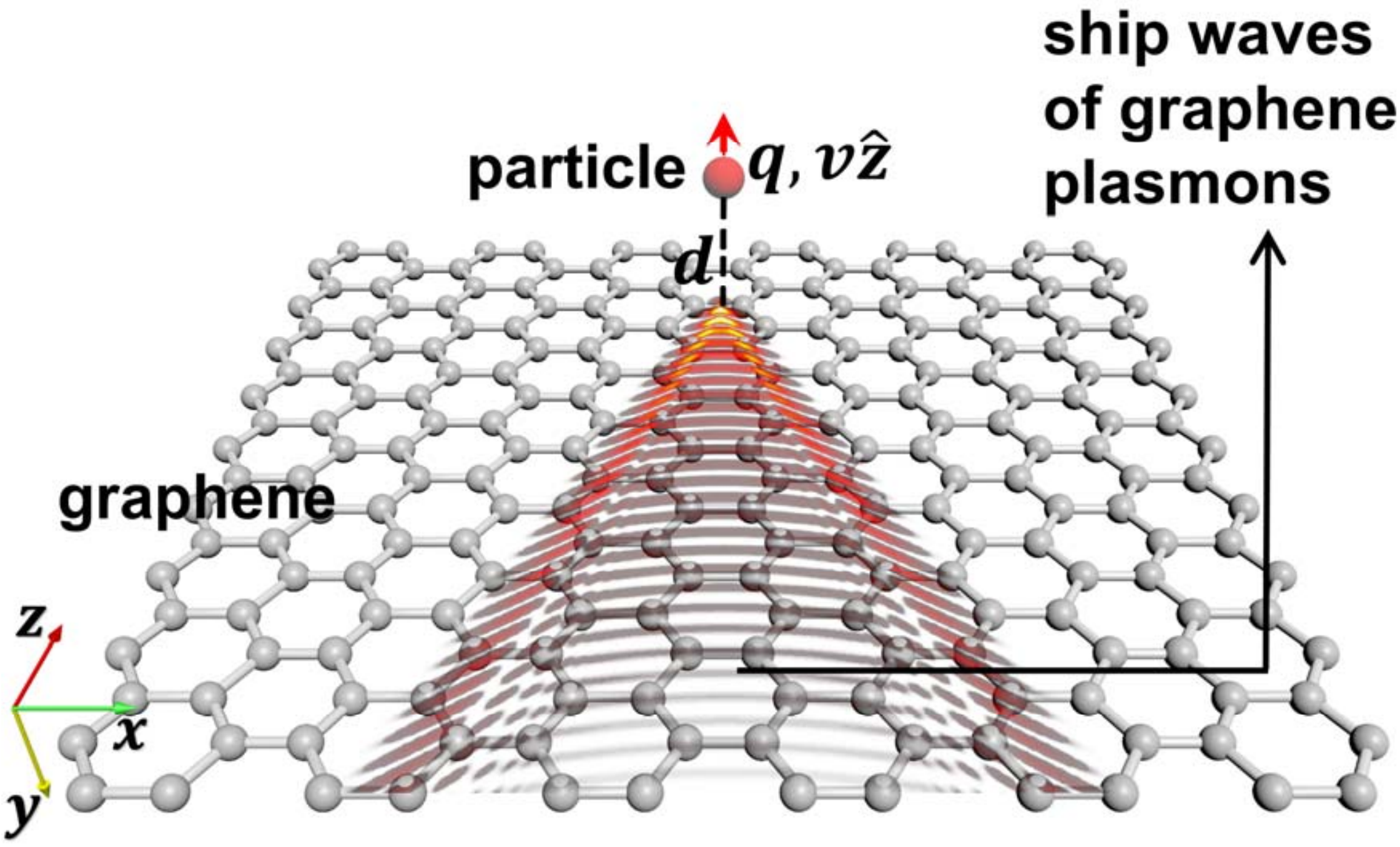

FIG. 1 (color online). The ship waves of graphene plasmons excited by a charged particle moving along the z-direction with constant velocity $v$. The distance between the graphene and the particle is $d$.

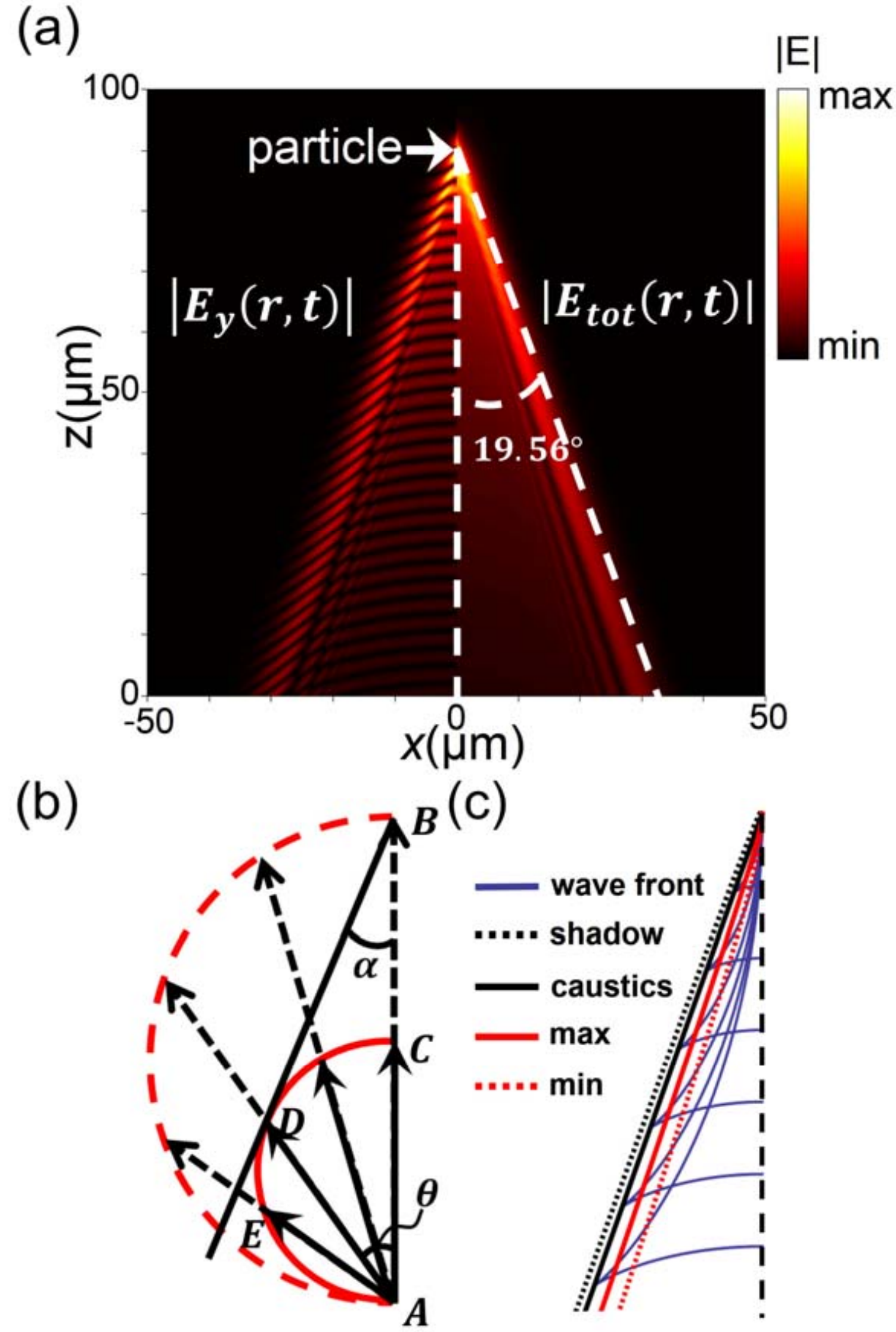

FIG. 2 (color online). The ship wave of graphene plasmons excited by a swift charged particle with velocity $v = 0.1c$. (a) The absolute value of electric field $|E_y(r,t)|$ (left) and the absolute value of the total electrical field $|E_{tot}(r,t)|$ (right) of the wave pattern. The arrow indicates the position of the paticle. (b) The Kelvin angle can be determined in a simple geometry. (c) The blue lines represent the wave fronts. The black dotted line is the boundary of caustic shadow. The black solid line is the caustic boundary. The red line and the red dotted line are the maximum and minimum amplitudes in the neighborhood of the caustics.

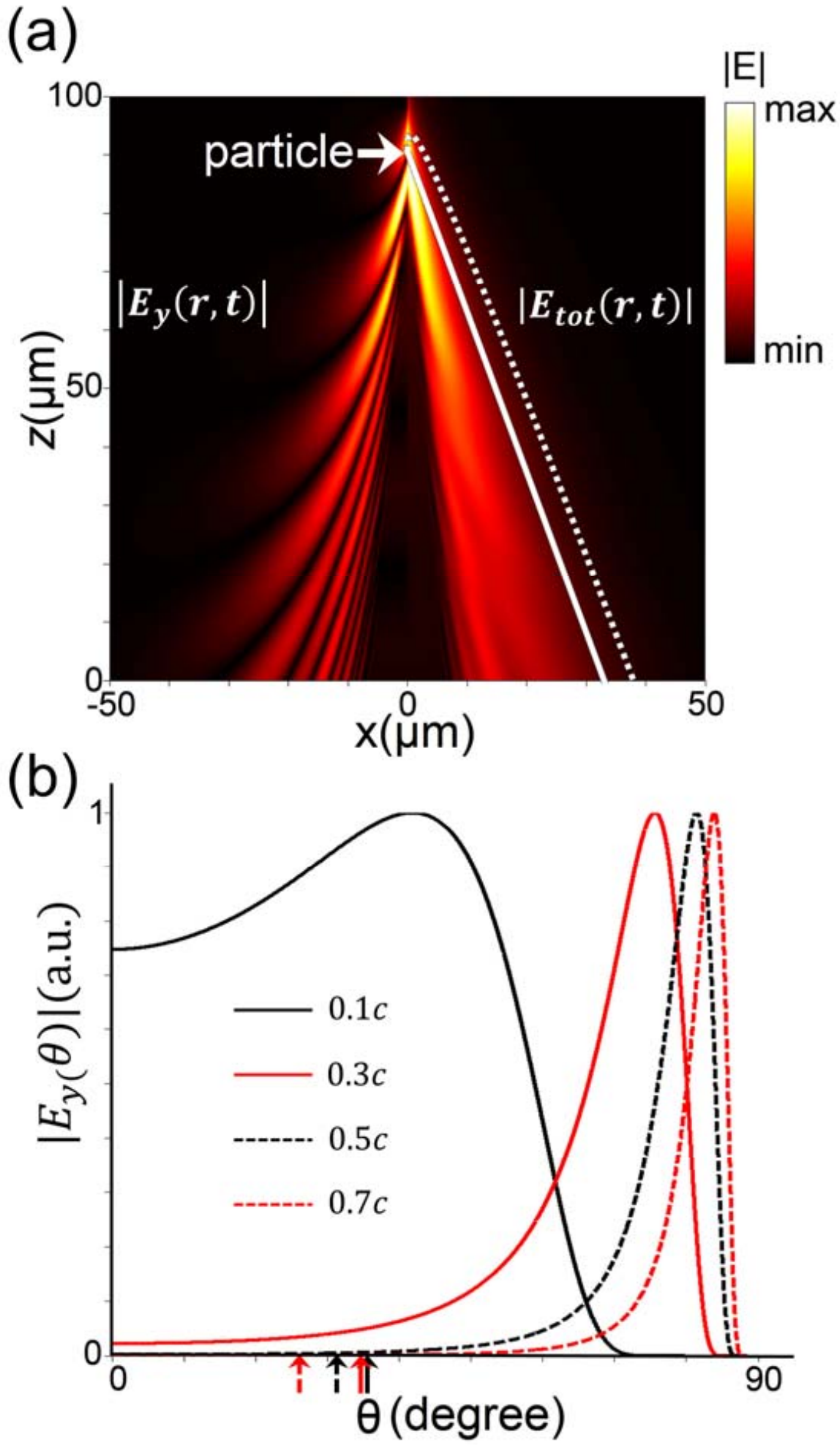

FIG. 3 (color online). The ship waves of graphene plasmons excited by a swift charged particle with velocity $v = 0.3c$. (a) The electric field $|E_y(r,t)|$ (left) and the total electrical field $|E_{tot}(r,t)|$ (right) of the plasmonic ship waves. The arrow indicates the position of the particle. The white solid line represents the caustics and the white dotted line is the boundary of the caustic shadow. (b) The normalized distribution of $|E_y|$ over propagation angle $\theta$. The velocity of the particle varies from $0.1c$, $0.3c$, $0.5c$ to $0.7c$. The arrows at the horizontal axis point to the angles at which the waves can reach the caustics.

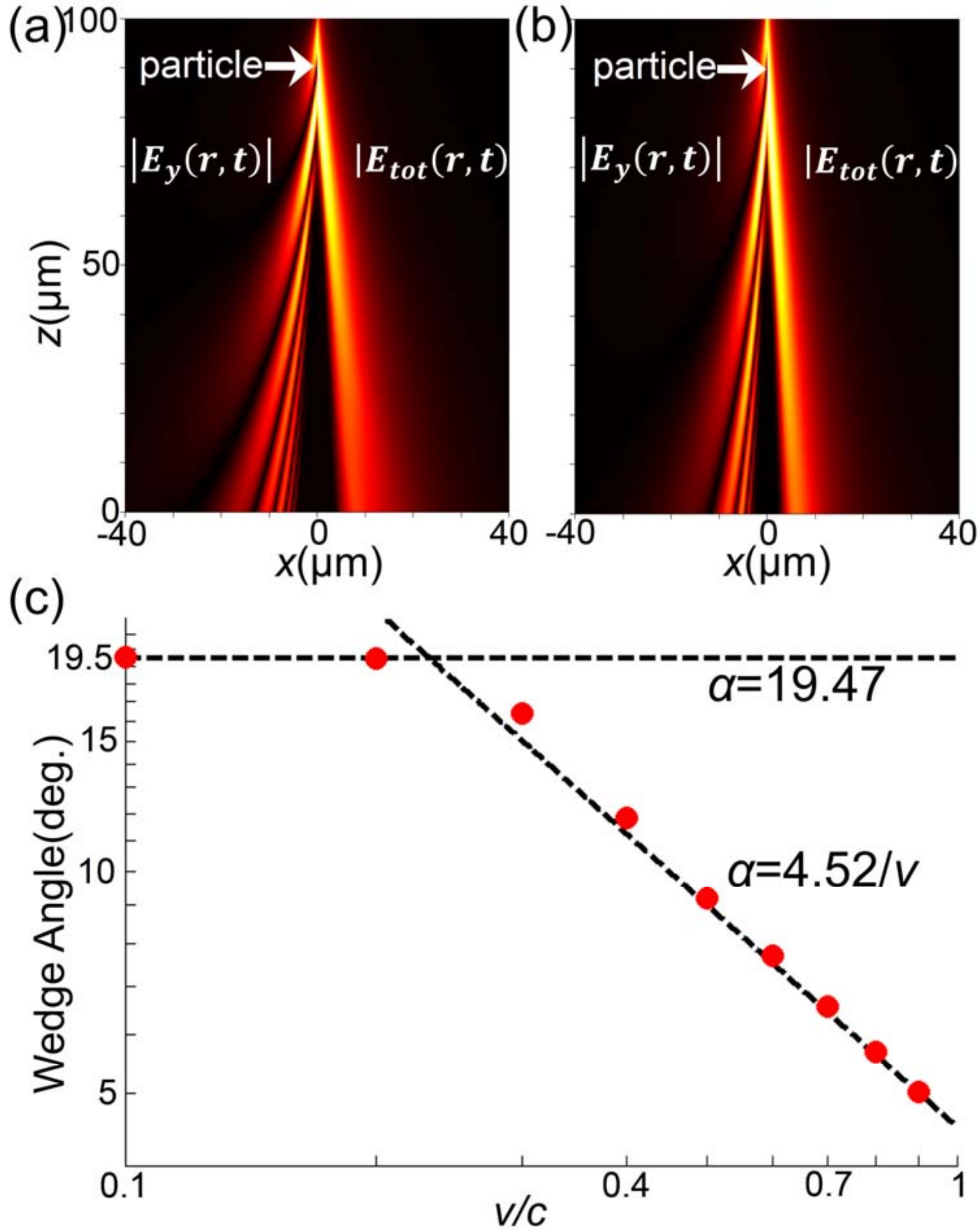

FIG. 4 (color online). The ship waves of graphene plasmons excited by a swift charged particle with velocity (a) $0.5c$, and (b) $0.7c$. (c) The dependence of effective semi-angle on the velocity of the particle. $\alpha = 19.47$ (degree) is the angle of the caustic boundary at zero velocity limitation. $\alpha = 4.52/v$ (degree) is the asymptote of the effective semi-angle at large particle velocity.